\documentclass[ reprint,  amsmath,amssymb,  aps, ]{revtex4}%
\usepackage{graphicx}
\usepackage{dcolumn}
\usepackage{bm}
\usepackage{amsmath}
\usepackage{amsfonts}
\usepackage{amssymb}%
\providecommand{\U}[1]{\protect\rule{.1in}{.1in}}
\begin{document}

\begin{abstract}

\end{abstract}

%

\preprint{APS/123-QED}%
%

\title{Neutrino
nuclear
response
and
photo
nuclear
reaction }%
%

%

\author{H. Ejiri}%
%
\email{ ejiri@rcnp.osaka-u.ac.jp}
\affiliation{Research Center for Nuclear Physics,
Osaka University, Ibaraki, Osaka, 567-0047, Japan\\
Nuclear Science, Czech Technical University, Prague, Czech Republic }%

\author{A.I. Titov}%

\affiliation{Joint
Institute of Nuclear Research, 141980, Dubna, Russia}

\author{M. Boswell}%
\affiliation{Los Alamos National Laboratory, Los Alamos, NM, USA}
\author{
A. Young}%
\affiliation{Department of Physics, North Carolina State University, Raleigh, NC, USA}

\date{\today}%

\begin{abstract}%
Photo nuclear reactions are shown to be used for studying neutrino/weak nuclear responses involved in astro-neutrino nuclear interactions and double beta decays. Charged current weak responses for ground and excited states are studied by using photo nuclear reactions through isobaric analog states of those states, while neutral current weak responses for excited states are studied by using photo nuclear reactions through the excited states. The weak interaction strengths are studied by measuring the cross sections of the photo nuclear reactions, and the spin and parity of the state are studied by measuring angular correlations of particles emitted from the photo nuclear reactions. Medium-energy polarized photons obtained from laser photons scattered off GeV electrons are very useful. Nuclear
responses studied by photo nuclear reactions are used to evaluate neutrino/weak nuclear responses, i.e. nuclear beta and
double beta matrix elements and neutrino nuclear interactions, and to verify theoretical calculations for them.\\



\end{abstract}%

\maketitle

\section{\label{sec:level1}Introduction}

Fundamental properties of neutrinos and astro-neutrino nuclear interactions are studied
by investigating nuclear double beta ($\beta \beta $) decays, nuclear inverse beta ($\beta $) decays and neutral current nuclear excitations. Here  nuclear weak responses (square of the nuclear matrix element) are crucial to study neutrino properties of particle and astro physics interests \cite{eji00,eji05,eji06}.

Neutrino-less double beta decays ($0\nu\beta \beta $) are most sensitive and realistic probes to study the Majorana properties of neutrinos, their absolute mass scales and the mass spectrum, the lepton-sector CP phases and others beyond the standard model. They are discussed in review papers and references therein \cite{eji05,eji06,avi08,eji09,eji10,ver12}. The 0$\nu \beta \beta $ transition rate via the $\nu-$mass process is given in terms of the effective $\nu $ mass $m_{\nu}$ as
\begin{equation}
T^{0\nu}=G^{0\nu }~(M^{0\nu })^2~(m_{\nu })^2,
\end{equation}
where $T^{0\nu}$ is the transition rate, $G^{0\nu }$ is the phase space factor, and $M^{0\nu}$ is the 0$\nu \beta \beta $ matrix element. Thus one needs the 0$\nu \beta \beta $ matrix element $M^{0\nu}$ to design the optimum detector and to extract the neutrino mass from $0\nu\beta \beta $ rate when it is observed.

The 0$\nu \beta \beta $ matrix element is expressed as $M^{0\nu}=\sum _k M_k^{\beta \beta}$, where $M_k^{\beta \beta}$ is the 0$\nu \beta \beta $ matrix elements  via th kth state in the intermediate nucleus . Then the single $\beta $ matrix elements of $M_k(\beta ^-)$ and $M_k(\beta ^+)$ for the $\beta $ transitions via the kth intermediate state can be used to help evaluate the $\beta \beta $ matrix element of $M_k^{\beta \beta}$ involved in $M^{0\nu }$. The 0$\nu \beta \beta $ decay is associated with medium momentum exchange of an intermediate virtual neutrino, and accordingly the intermediate states involved are the ground and excited states with angular momentum of $l \approx $1-3.

Astro-neutrino charged current (CC) interactions  are studied by measuring the inverse $\beta ^{\pm}$ decays induced by the neutrinos, and then the $\beta $ matrix element of $M(\beta ^{\pm})$ are needed to study the interactions. Similarly neutral current (NC) studies require the NC matrix element.

Accurate calculations of $\beta $ and  $\beta \beta $ matrix elements are hard since they are sensitive to nuclear spin isospin correlations, nuclear medium effects, and nuclear structures, as discussed in the review articles \cite{eji00,eji05,ver12,suh98,sim11}. Accordingly, experimental studies of them are of great interest \cite{eji00,eji05,eji06,eji09,ver12}.

The nuclear weak responses (matrix elements) are studied experimentally by using neutrino/muon probes with weak interactions, photon probes with electro magnetic (EM) interactions, and nuclear probes with strong nuclear interactions \cite{eji00,eji06,eji10,ver12}, as shown in Fig.1.

Neutrino beams could be useful if intense neutrino beams and multi-ton scale detectors might be available. The neutrinos from SNS and J-PARK are of great interest \cite{eji05,eji03,avi00}.
Recently $\mu $ capture reactions are shown to be used to study $\beta ^+$ strengths \cite{eji06,suh06,eji13}.

Charge exchange reactions by using nuclear probes have been extensively used for evaluating CC $\beta ^{\pm}$ responses. The high energy-resolution ($^3$He, t) reaction at RCNP is very powerful for
studying CC $\beta ^-$ responses \cite{ver12,aki97,thi12,thi12a}. CC $\beta ^+$ response studies by (t,$^3$He) reactions require radio-active t beams \cite{gue11}, and those by (d,$^2$He) reactions need a big spectrometer for $^2$He$\rightarrow $2p \cite{doh08}.


\begin{figure}[t]
\includegraphics[scale=0.25]{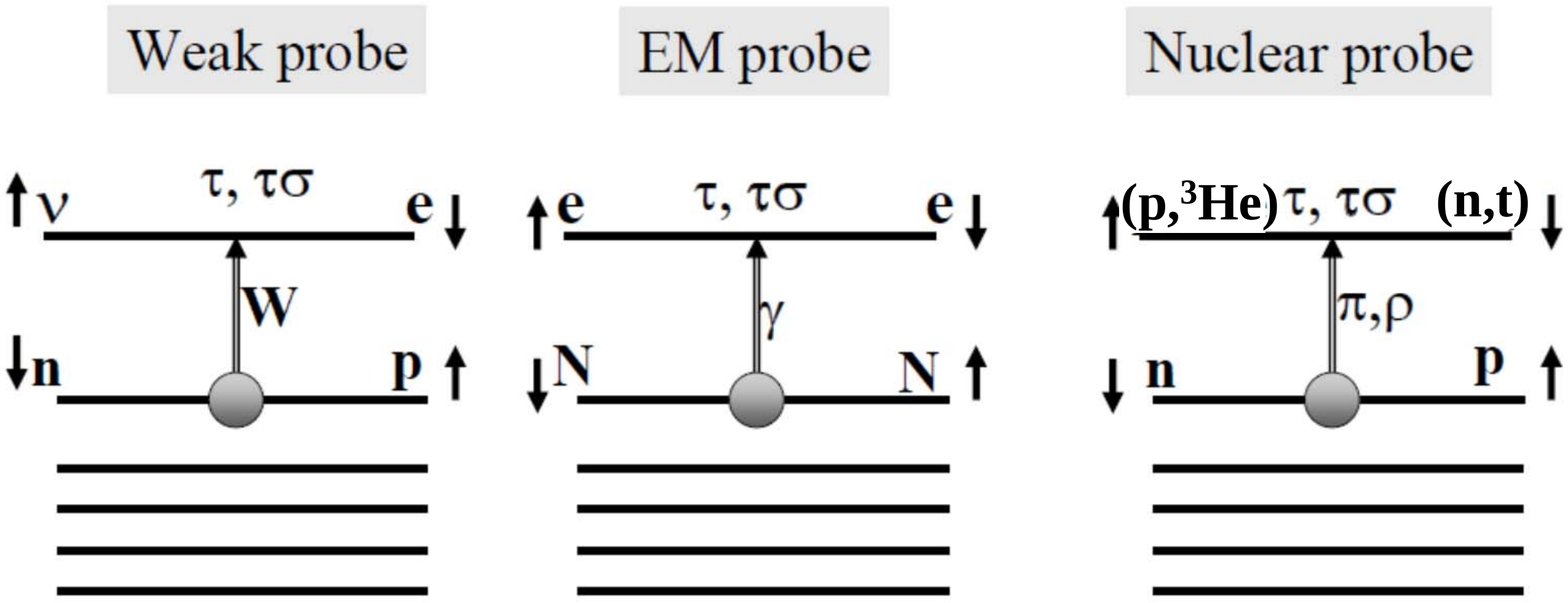}
\caption{Nuclear spin ($\sigma )$ and isospin ($\tau $) responses for CC weak interactions and their studies by  neutrino ($\nu $) probes via weak interaction, $\gamma $ probes via EM interaction and by nuclear probes via strong nuclear interaction \cite{eji00,eji05}.}%
\label{fig:epsart}%
\end{figure}

The present work aims at reporting possible photo nuclear reactions by using high quality EM photon probes for CC $\beta ^+$ and NC responses. In fact, weak and EM responses have similar spin isospin operators, and thus the EM responses are used to evaluate the weak ones, and vise versa, as used for transitions via isobaric analogue state (IAS) \cite{fuj67,eji68,eji68a} and also for electron scatterings and others \cite{boh75,don75,eji78}. The use of EM probes for $\beta \beta $ response studies is discussed in reviews \cite{eji03,eji10}. Unique features of EM photon probes are as follow.

i. EM interactions involved in photo nuclear reactions are well known. The dominant lowest multi-pole transition is well evaluated within the long wave-length approximation in the present excitation region.

ii. NC and CC $\beta ^+$ responses for nuclear ground and excited states are studied by measuring photo nuclear excitations of them and those of the IASs, respectively.

iii. Vector ($\tau$ isospin) and axial vector ($\sigma \tau$ spin isospin) responses are studied by measuring electric (E) and magnetic (M) photo nuclear excitations, respectively.

iv. Spin and parity of the state are determined by measuring angular correlations of photo nuclear reactions with polarized photons.

v. EM interactions are simple and photon probes are not distorted in nuclei, while nuclear interactions involved in nuclear probes are rather complicated and nuclear beams are distorted by nuclear potential.

vi. High energy-resolution intense photon beams with large polarization are obtained from polarized laser photons scattered off GeV electrons.

\section{WEAK RESPONSES STUDIED By PHOTO NUCLEAR REACTIONS}

Weak and EM transitions to be discussed in the present paper are the vector and EL(E1,E2) transitions with natural parity $J^P =1^-,2^+$ and the axial-vector and ML(M1,M2) transitions with unnatural parity $J^{\pi} =1^+,2^-$. The weak vector and axial vector transition operators are expressed as \cite{eji00,eji78,boh75}

\begin{equation}
T(VL)= g_V \tau ^i r^LY_L  ~~~~~~~~L =1, 2,
\end{equation}
\begin{equation}
T(AV L)= g_A \tau ^i [\sigma \times r^{L-1}Y_{L-1}]_L ~~~~~~~~L =1, 2,
\end{equation}
where $g_V$ and $g_A$ are the vector and axial vector weak coupling constants, and $\tau ^3$ and $\tau ^{\pm}$ are NC and CC isospin operators. The EM transition operators corresponding to the VL and AVL weak ones are expressed as \cite{eji00,eji78,boh75}
\begin{equation}
T (EL)= g_{EL} r^LY_L,
\end{equation}
\begin{equation}
T(ML)= g_S[\sigma \times r^{L-1}Y_{L-1}]_L +g_L[j \times r^{L-1}Y_{L-1}]_L,
\end{equation}
\begin{equation}
g_S = \frac{e\hbar}{2Mc}[L(2L+1)]^{1/2}~[\frac{g_s}{2}-\frac{g_l}{L+1}]
\end{equation}
\begin{equation}
g_L = \frac{2g_l}{L+1},
\end{equation}
where $g_{EL},g_S$ and $g_l$ are the EM coupling constant (effective charge), the spin $g$ factor and the orbital $g$ factor, respectively. In case of spin stretched ML transitions of $J\rightarrow J\pm L$, the second term of the $T(ML)$ vanishes, and the transition operator is given by the first term of eq.(5) \cite{eji78,boh75}.
Then experimental studies of EL and Ml transition rates are used to evaluate the analogous vector and axial vector weak responses.

The weak axial vector transitions of $T(AV J)=g_A \tau ^i[\sigma \times rY_1]_J$ with $J^{\pi} =0^-,1^-, 2^-$ are of the same order of the magnitude as $T(V 1)$. However, the corresponding EM transitions are 2-3 orders of magnitudes weaker than $T(E1)$. Then the 1$^-$ contribution is negligible compared with the E1 transition. There is no $0^+\rightarrow 0^- \gamma$ transition. Thus we discuss only $J^{\pi}=2^-$.

The EM coupling constant depends on the isospin $z$ component, namely on the proton or neutron. Using the isospin operator $\tau ^3$ and $\tau ^0$ = 1, the coupling constant is expressed as
\begin{equation}
g_i = \frac{g_i(V)}{2}\tau^3  + \frac{g_i(S)}{2}\tau ^0,
\end{equation}
where $g_i$ with i=E, s, and l are the electric, the magnetic spin, and the magnetic orbital coupling constants, and
$g_i(V )$ and $g_i(S)$ are the corresponding isovector and isoscalar coupling constants. They are written by using the neutron and proton coupling constants as $g_i(V )= g_i(n) - g_i(p)$ and $g_i(S)= g_i(n)+ g_i(p)$.

The EM transition matrix element includes both the isovector and isoscalar components, and they are modified differently by nuclear spin isospin correlations and nuclear medium effects. Consequently, the isovector weak matrix element is not exactly the same as  the corresponding EM matrix element. Nevertheless, the measured EM matrix element helps evaluate the weak matrix element and check/confirm theoretical calculations of the weak matrix element. IAS provides a unique opportunity to select exclusively the isovector component of the EM transition, which is analogous to the $\beta ^+$ transition. So, we discusses mainly IAS $\gamma $ transitions in the present paper.

\begin{figure}[h]
\includegraphics[width=0.3\textwidth]{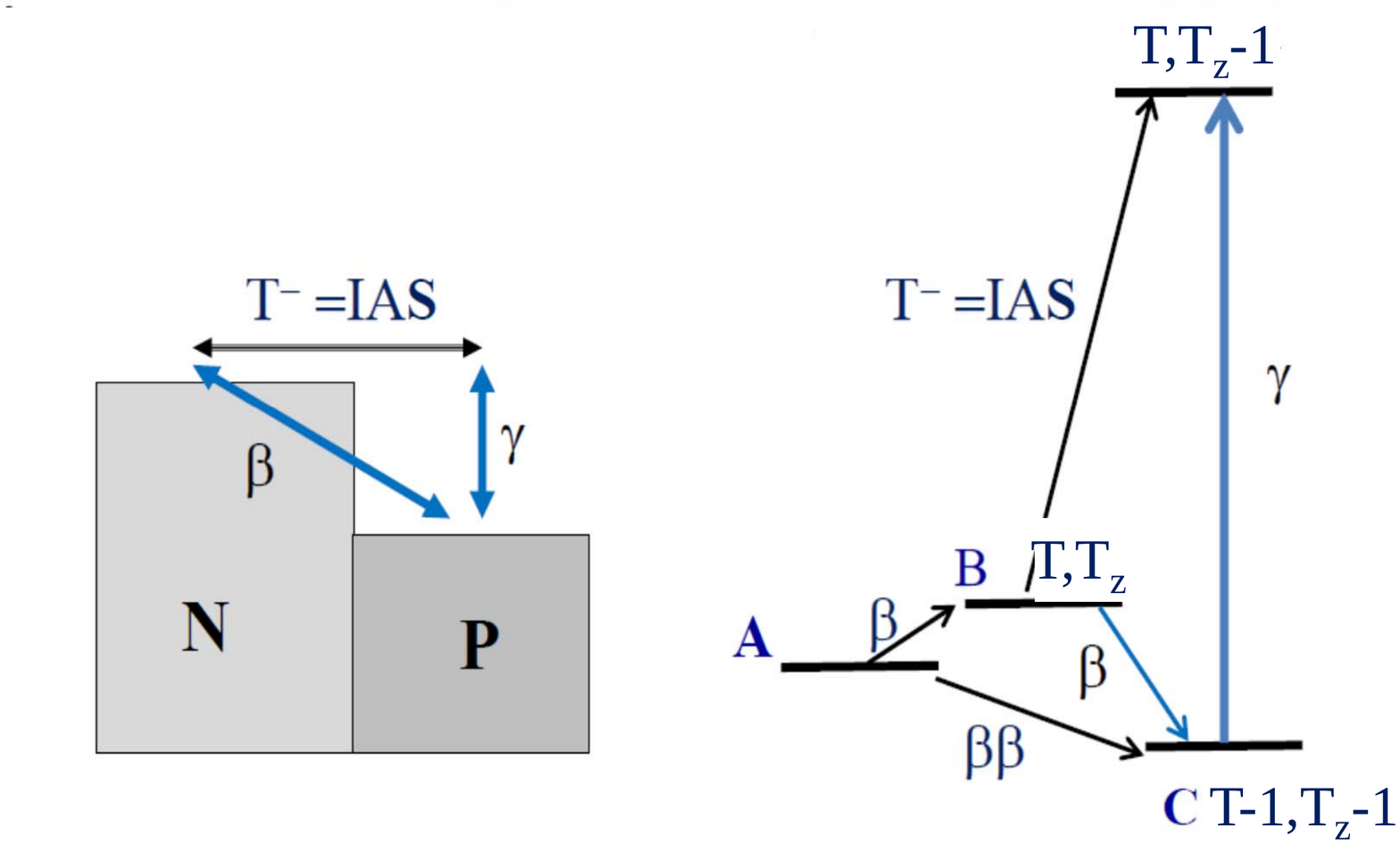}
\caption{Level and transition schemes of single ƒÀ and ƒÁ transitions via IAS. T$^-$ is the isospin lowering operator to transfer a state to IAS \cite{fuj67,eji68,eji68a}. $T_z$(=$T$) is the isospin $z$ component}.%
\label{fig:epsart}%
\end{figure}

\section{PHOTO NUCLEAR REACTIONS VIA ISOBARIC STATES}

CC $\beta ^+$ responses are studied by photo nuclear reactions through IAS, as shown by the $\gamma $ decays from IAS \cite{fuj67,eji68,eji68a}. The $\beta $  and $\gamma $ matrix elements are related as
\begin{equation}
<f|g m^{\beta }|i>\approx \frac{g}{e}K<f|e m^{\gamma }|IAS >,
\end{equation}
where $g m^{\beta }$ and $e m^{\gamma }$ are analogous $\beta $ and $\gamma $ transition operators with $g$ and $e$ being the weak and EM coupling constants, $K=(2T)^{1/2}$ is the normalization constant, and $|IAS >= K^{-1}T^-|i>$ with $T^-$ being the isospin lowering operator.

The relation between the weak and EM matrix elements for IAS transitions is based on the rotation ($T^-$) symmetry in the isospin space.   Then one can obtain the
$\beta $ matrix element for $ |f>\rightarrow |i>$ by observing the analogous $\gamma $ absorption for $|f>\rightarrow |IAS>$. Here $|f>$ and $|i>$ are the final and intermediate states in the $\beta \beta $ decay, and the initial and final nuclei in the anti-neutrino CC nuclear interaction. This is a noble method used extensively for (p,$\gamma $) reactions through IAS on various nuclei as discussed in \cite{eji89}. The $\beta,~ \beta \beta $, and analogous $\gamma $ transitions are schematically shown in Fig.2.

In medium and heavy nuclei, IAS is observed as an isobaric analogue resonance (IAR) in the medium excitation region.
The photo nuclear cross section via the IAR with $J^{\pi}$ is expressed as
\begin{equation}
\sigma (\gamma, n)= \frac{S(2J+1) \pi}{k_{\gamma }^2}\frac{\Gamma _{\gamma } \Gamma _n}{(E-E_R)^2+\Gamma _t^2/4},
\end{equation}
where $\Gamma _{\gamma },\Gamma_t$, and $\Gamma _n$ are the $\gamma $ capture width, the total width and the neutron decay width, $S$ is the spin factor, and $k_{\gamma }$ is the incident photon momentum.

The integrated photo nuclear cross section is given by
\begin{equation}
\int \sigma (\gamma, n)dE= \frac{S(2J+1) 2\pi ^2}{k_{\gamma }^2}\frac{\Gamma _{\gamma } \Gamma _n}{\Gamma _t}.
\end{equation}
.

IAR at the excitation region well above the particle threshold energy decays by emitting mostly neutrons since proton decays are suppressed by Coulomb barrier. Then one gets $\Gamma _t\approx \Gamma _n$, where $\Gamma _n$ is the sum of the neutron decay widths to all states. In case of the $0^+ \rightarrow 1^{\pm}$ excitation in even even nuclei, the $\Gamma _{\gamma }$ width is obtained from the measured integrated cross section $\int  \sigma(\gamma, n)dE = \pi^2k_{\gamma }^{-2}\Gamma _{\gamma }$, where $\sigma(\gamma , n)$ is the sum of the $(\gamma ,n)$ cross sections for neutron decays to all final states.

Now we discuss photo nuclear cross sections for E1 and M1 photo excitations on even even nuclei with $J^{\pi} = 0^+$.  The reduced widths of $\Gamma _{\gamma }(E1)$ and $\Gamma _{\gamma }
(M1)$ are expressed, respectively, in terms of the E1 and M1 matrix elements of $M(E1)efm$ and $M(M1)e/(2Mc)$ and the excitation(photon) energy $E_{\gamma }$ in unit of MeV as \cite{boh75}
\begin{equation}
\Gamma _{\gamma}(E1) = 1.59\times10^{15} ~ E^3 M(E1)^2 /sec,
\end{equation}
\begin{equation}
\Gamma _{\gamma }(M1) = 1.76\times10^{13} ~ E^3 M(M1)^2 /sec.
\end{equation}
The $M(E1)$ and $M(M1)$ for IAR are expressed as $M(E1)=M_1(E1)/K$ and
$M(M1)=M_1(M1)/K$, where $M_1(E1)$ and $M_1(M1)$ are the corresponding isovestor $\gamma $ matrix elements.

The matrix elements for typical E1 and M1 excitations on medium heavy nuclei in the mass region of $A$=100, $T$=8, $K=4$ are $M_1(E1)\approx 0.125$ \cite{eji78} and $M_1(M1)\approx 1.15$ \cite{eji78}. Using these values, for example, one gets $M(E1)\approx 0.03$ and $M(M1)\approx 0.28$ for the photon absorption into IAS. The IAR excitation energy is around E=8 MeV in this mass region. Then the cross sections are obtained as

\begin{equation}
\int \sigma (\gamma ,n)dE =2.9\times 10^{-3} MeV fm^{2}~~~ E1
\end{equation}
\begin{equation}
\int \sigma (\gamma ,n)dE =2.7\times 10^{-3}  MeV fm^{2}~~~M1 .
\end{equation}
Then the counting rates with a typical target of 10 g/cm$^2$ are $Y(E1)$=1.7$\times 10^{-6} \epsilon N_{\gamma }$/sec and $Y(M1)$=1.6 $\times 10^{-6}\epsilon N_{\gamma }$/sec, where $N_{\gamma }$ is the number of photons per sec per MeV and $\epsilon $ is the detection efficiency. Thus experimental studies of these photo nuclear reactions are quite realistic by using medium energy photons with $N_{\gamma } \approx $10$^{8-9}$/(MeV sec).

In heavy nuclei, IAR shows up as a sharp resonance on the top of the E1 giant resonance (GR). Then the E1 photo nuclear reaction is given by \cite{eji68a},
\begin{equation}
\frac{d\sigma (\gamma ,n)}{d\Omega}= k[A_I^2 + A_G^2 + 2Re(A_I A_G~e^{i\phi})],
\end{equation}
where $A_I$ and $A_G$ are the IAR and GR amplitudes and  the third term is the interference term with $\phi$ being the relative phase. It is noted that the phase of the matrix element can be determined from the interference pattern.

The two neutrino $\beta \beta $ matrix element is expressed by the sum of the products of the single $\beta ^-$ and $\beta ^+$ matrix elements via low-lying  intermediate $1^+$ states \cite{eji09}. Similarly, the 0$\nu\beta \beta $ matrix element may be given approximately by those of the single $\beta ^-$ and $\beta ^+$ matrix elements via low-lying intermediate  states with $J^{\pi} =0^{\pm}, 1^{\pm}, 2^{\pm}, 3^{\pm}$ and so on. In cases of medium energy neutrino interactions $0^{\pm}, 1^{\pm}, 2^{\pm}$ states are involved .
The single $\beta ^+$ matrix elements for the 1$^{\pm}$ states are studied by measuring photo nuclear reactions via the IASs of the $1^{\pm}$ states.

\section{ANGULAR CORRELATIONS OF PHOTO NUCLEAR REACTIONS}

Nuclear states excited by medium energy photons decay by emitting mostly neutrons.  The spin and parity of the excited state are identified by measuring angular distributions of the emitted particles with respect to the photon beam and its polarization.

Let us consider the photo nuclear reaction on an even-even nucleus by using linearly polarized photons $\vec{\gamma }$,
\begin{equation}
\vec{\gamma }+N(0^+) \rightarrow N^*(J^P)\rightarrow N_r(J_r^P) + n,
\end{equation}
where N, N$^*$, and N$_r$ are the target nucleus in the ground state, the excited nucleus and
the outgoing recoil nucleus, respectively, with corresponding spin-parities of 0$^+, J^P$ and $J_r^P
$, and n is the outgoing neutron. We calculate the azimuthal and polar angular
distributions of the outgoing neutron. We choose z axis along incoming photon momentum,
and x axis along the photon polarization.

The angular distribution is defined as
\begin{equation}
W(\theta,\phi) = \sum _{\sigma M_r} |A^{S}(JJ_r\sigma M_r\theta \phi)|^2,
\end{equation}
where $A^{S}$ is the amplitude for excitation of the target nucleus to the excited state and
 subsequent decay to the recoil nucleus and
the neutron. The subscript S stands for natural (S=N) or unnatural (S=U) spin-parity excitation.
$M_r$ and $\sigma $ are the spin projections of the recoil nucleus and the neutron,
respectively, and $\theta $ and $\phi $ are the polar and the azimuthal angles of the neutron momentum, respectively.
The angular distribution is normalized as
\begin{equation}
\int W(\theta,\phi)dcos\theta d\phi = 1.
\end{equation}
The transition amplitude is given as
\begin{equation}
A^S(JJ_r\sigma M_r\theta \phi)=\sum _{lm}a_lC^S(JJ_rlm\sigma M_r)Y_{lm}(\theta \phi),
\end{equation}
where $a_l$ is the partial amplitude of the neutron decay with the orbital angular momentum $l$, and
 $C^S$ is the coefficient for the angular momentum couplings as written in terms of the Clebsch Gordan coefficients for the coupling of the angular momenta involved in the reaction.

Let us consider photo nuclear reactions on two typical nuclei of $^{76}$Se and $^{100}$Mo. The photo excitations discussed are the EM transitions to the IASs of $^{76}$As and $^{100}$Tc. Thus they are used to evaluate the analogous weak transitions to $^{76}$As and $^{100}$Tc with the same multi-pole and spin-parity.
\begin{equation}
\vec {\gamma} +^{76}\rm{Se} \rightarrow ^{76}\rm{Se}(J^p)\rightarrow ^{75}\rm{Se}(\frac{3}{2}^-) + n,
\end{equation}
\begin{equation}
\vec{\gamma }+^{100}\rm{Mo} \rightarrow ^{100}\rm{Mo}(J^p)\rightarrow ^{99}\rm{Mo}(\frac{1}{2}^+) + n.
\end{equation}

The odd neutron of the recoil nucleus $^{75}$Se is in the orbital p state, while that of the
recoil nucleus $^{99}$Mo is in the s state. This
difference is important for the angular distribution of the outgoing neutron.
We will consider the azimuthal angular distribution at the fixed polar angle $\theta = \pi$/2 and the polar angular
distribution at the fixed azimuthal angle $\phi = \pi$/2.\\

i. Natural parity excitations

The angular distributions for natural parity 1$^-$ excitations are presented in Fig. 3. In case of Se, the possible orbital configurations of the outgoing neutron from the $1^-$ state to the (3/2)$^-$ ground state in $^{75}$Se
are s ($l$=0) and d ($l$=2). The first case leads to the isotropic distribution, while the second one gives the
anisotropic distribution, determined by $Y_2$-spherical harmonics. We introduce a variable $x$
defined as the relative probability of the d neutron emission. It is proportional to $a_{l=2}^2$. We use $x$ as a parameter that varies from 0 to 1.0. The distributions with $x$ = 0, 0.2, 0.4, 0.6, 0.8, and 1.0 are shown in Fig.3. In case of Mo, only p ($l$=1) neutron emission is
allowed from the 1$^-$ state to the 0$^+$ ground state in $^{99}$Mo. The distribution is close (but not exactly) to sin$^2\phi$.

\begin{figure}[t]
\includegraphics[width=0.3\textwidth]{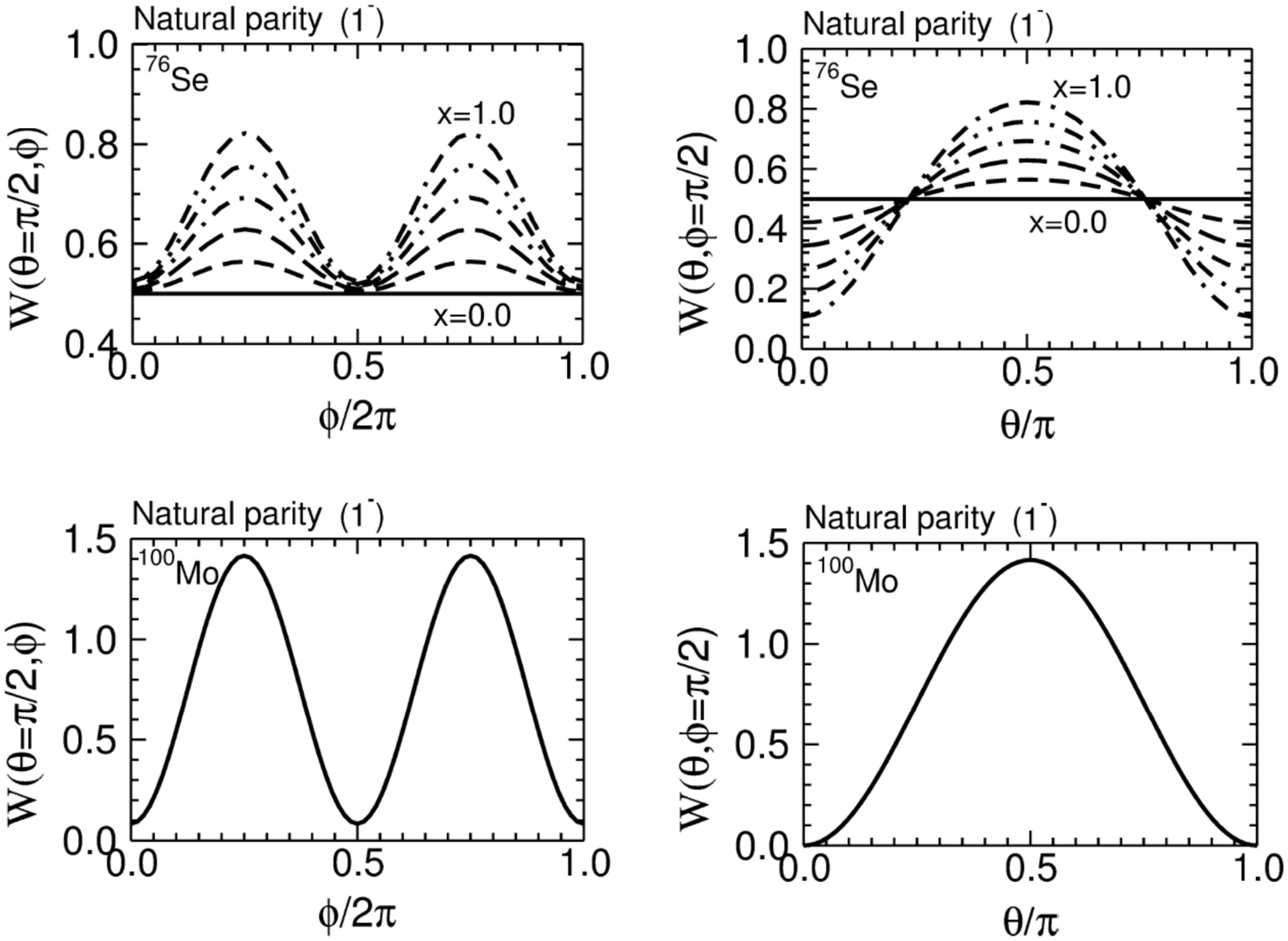}
\caption{Top: Azimuthal (left) and polar (right) angular distributions (relative) of the neutron from the 1$^-$ photo nuclear
excitation on $^{76}$Se with x being the fraction of the d configuration (see text). Bottom: The angular distributions for $^{100}$Mo.}%
\label{fig:epsart}%
\end{figure}

\begin{figure}[h]
\includegraphics[width=0.3\textwidth]{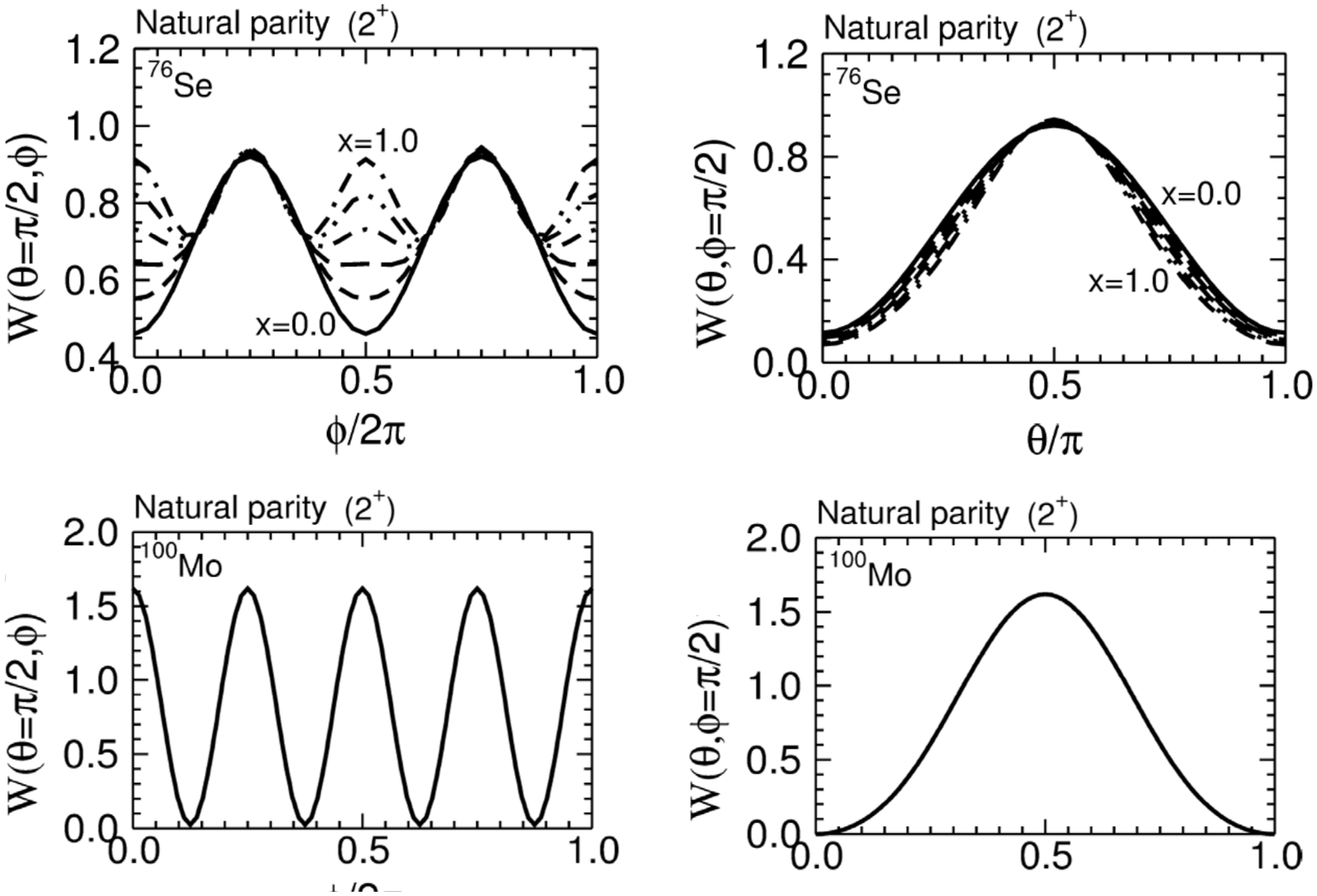}
\caption{Top: Azimuthal (left) and polar (right) angular distributions (relative) of the neutron from the 2$^+$ photo nuclear
excitation on $^{76}$Se with x being the fraction of the f configuration (see text). Bottom: The angular distributions for  $^{100}$Mo.}%
\label{fig:epsart}
\end{figure}

The angular distributions for natural parity 2$^+$ excitations are shown in Fig. 4.
In case of Se, the possible orbital configurations of the outgoing neutron to the (3/2)$^-$ ground state in $^{75}$Se are p and f ($l$=3). Here $x$ is the relative probability of the f neutron. Then, the angular distribution is defined by $Y_1$ and $Y_3$ spherical harmonics. In case of Mo nucleus, only the d configuration is allowed, which results in superposition of $Y_2^2$ harmonics.\\

ii. Unnatural parity states.

The angular distributions for unnatural parity 1$^+$ excitations are presented in Fig. 5.
In case of Se, the possible orbital configurations of the outgoing neutron to the (3/2)$^-$ ground state in $^{75}$Se are p and f, which result in anisotropic $\phi$ dependence with maximum at $\phi = \pi/2$ and $3\pi/2$. In Fig. 5 the variable $x$ denotes the relative contribution of the f orbital neutron.
The distributions for $x$=0 are not quite different from those for
the natural parity 1$^-$ state. Then, measurements of the neutron decays to other excited
states can be also used to identify the spin-parity of the state.
In case of Mo, the s and d configurations contribute.  Then
one gets the anisotropic distribution by interplay of $Y_0^2$ and $Y_2^2$ harmonics, depending on the relative weight $x$ of the d neutron.

\begin{figure}[h]
\includegraphics[scale=0.2]{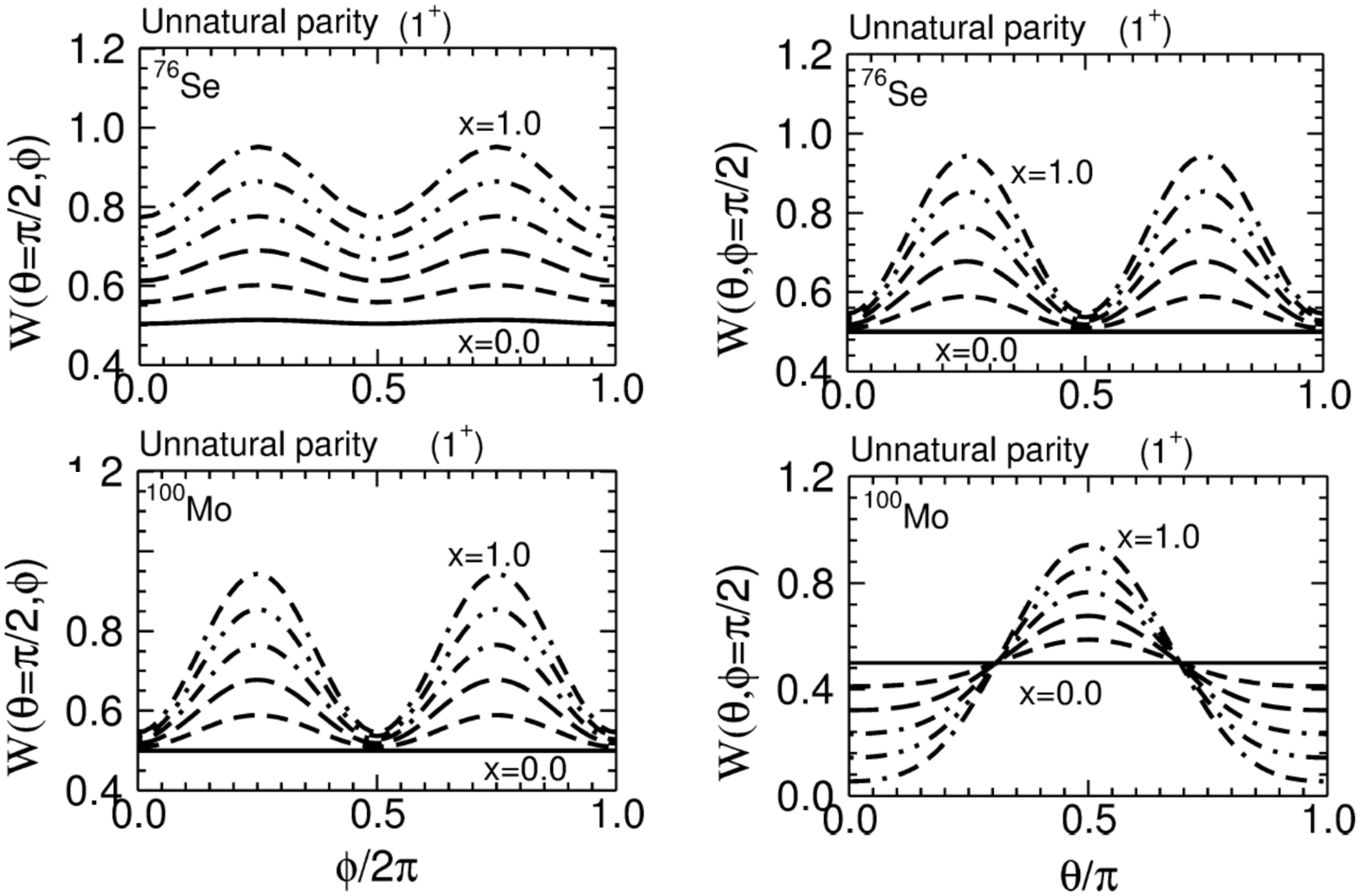}
\caption{Top: Azimuthal (left) and polar (right) angular distributions (relative) of the neutron from the 1$^+$ photo nuclear
excitation on $^{76}$Se with x being the fraction of the f configuration (see text). Bottom: The angular distributions for $^{100}$Mo with x being the fraction of the d configuration (see text).}%
\label{fig:epsart}%
\end{figure}
\begin{figure}[h]
\includegraphics[scale=0.2]{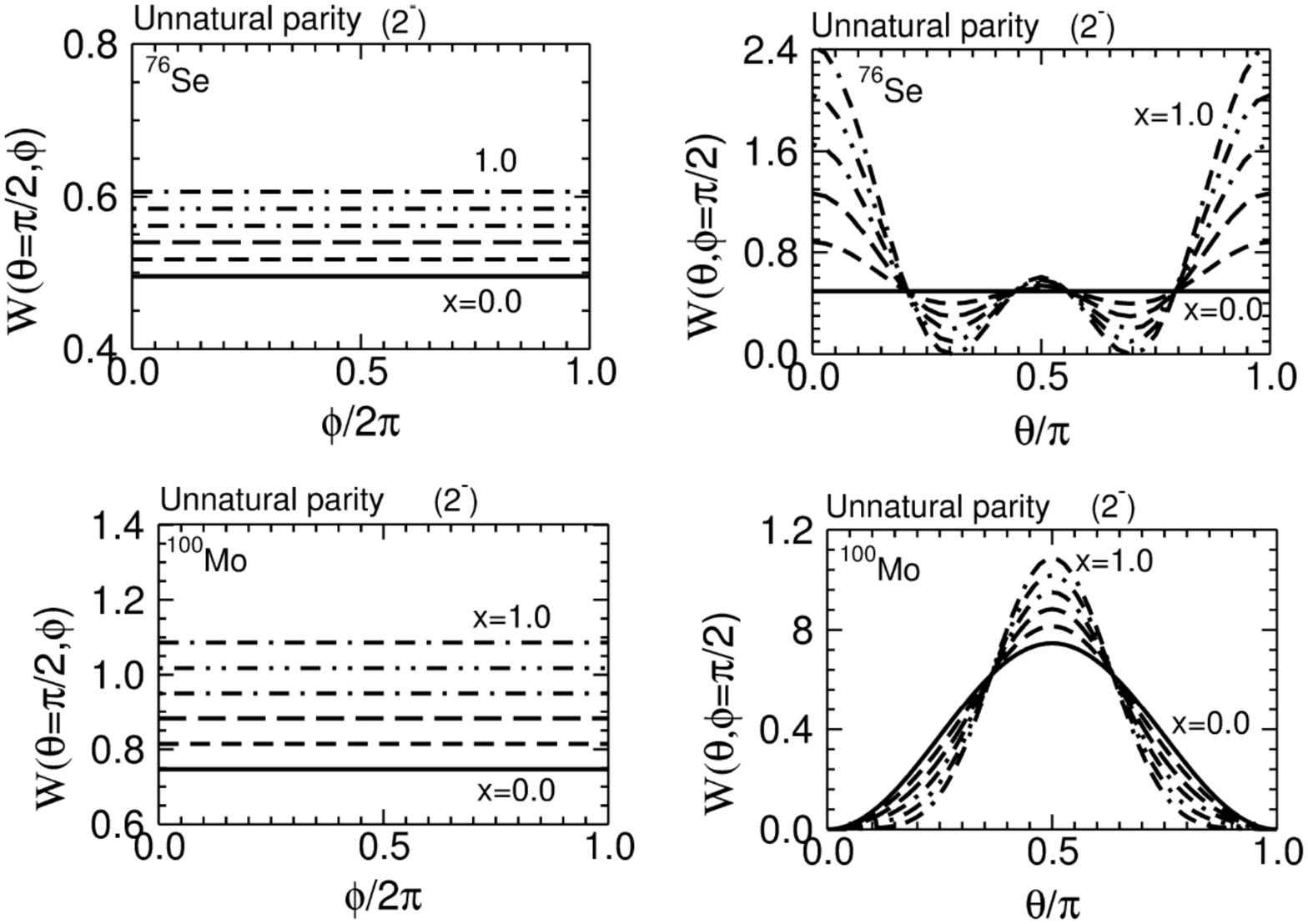}
\caption{Top: Azimuthal (left) and polar (right) angular distributions (relative) of the neutron from the 2$^-$ photo nuclear
excitation on $^{76}$Se with x being the fraction of the d configuration (see text). Bottom: The angular distributions for  $^{100}$Mo with x being the fraction of the f configuration (see text).}%
\label{fig:epsart}%
\end{figure}

Angular distributions of neutrons from an unnatural parity 2$^-$ state is shown in Fig. 6.
In case of Se, s and d orbital configurations are involved. Here $x$ is the relative probability
of the d neutron. The angular distributions are given by interplay of
$Y_0^2$ and $Y_2^2$ for s and d, respectively.  The first one gives isotropic distribution.
 In the second case, the sum over the Mr and spin projections results in the isotropic $\phi $ dependence. For the Mo nucleus,
the p and f configurations give the angular distribution defined by an interplay of $Y^2_1$ and $Y^2_3$ harmonics.
 Here $x$ is the relative probability of the f neutron. The sum over the neutron spin projection results in anisotropic $\phi$-dependence.

\section{REMARKS}
The present report shows by theoretical considerations and calculations that NC and CC $\beta ^+$ weak
responses are studied by using by photo
nuclear excitations. The EM coupling constant given in eq.(5) is mainly the spin ($g_s$) component in case of the isovector IAS transition as the axial-vector weak coupling constant, and they are renormalized similarly by spin isospin correlations. Thus EM IAS matrix elements give single $\beta ^+$ matrix elements associated with the $0\nu \beta \beta $ matrix element.
Actually, the renormalization effects on weak and EM transitions are in general not exactly the same, and one needs careful considerations on the possible state dependence in cases of EM transitions to different (non-IAS) states.

The $\beta $ transition operator of the first forbidden $\beta $ decays with $\Delta J$ = 1  includes three terms
of $T(rY1), T(\alpha)$ and $T(\sigma \times rY_1)_1$. Then the term $T(rY1)$ is derived from the analogous
E1 transition, and the term $T(\alpha)$ relative to $T(rY_1)$ is evaluated from the CVC theory.
Therefore, the spin matrix element $M(\sigma \times rY_1)_1$ can be deduced if both the $\beta $ and
 $\gamma $ transition rates are measured \cite{eji68,eji68a}.

Experimental studies of E1 and M1 photo nuclear reactions  are quite realistic, while studies of E2 excitations require
very intense photon beams. M2 excitations could hardly be realistic because of very small
cross sections. Inelastic electron scatterings for them are interesting.

High Intensity
$\gamma $-ray Source (HIGS) is very attractive. The intense $\gamma $ rays with E=2-70 MeV,
$\Delta E/E \approx 1\%$, and $I_{\gamma } \approx 10^7$/MeV/s are obtained by intra-cavity
Compton back-scattering of FEL photons off 1.2 GeV electrons at Duke Storage Ring \cite{wel09}.
Possible studies of photo nuclear excitations of IAS in $^{76}$Se were discussed \cite{bos13}.
The photon intensity will be increased by 2 orders of magnitude in future. CLS aims at intense photons
from CO$_2$ lasers scattered off the 3 GeV electrons in the CLS ring \cite{szp13}. Laser-backscattered
 source at NewSUBARU provides $\gamma $ rays with $E$=17-40 MeV, $\Delta E
/E$=2$\%$ and $I_{\gamma }\approx 10^7$/s \cite{miy12}. These photons are promising photon probes for the present photo nuclear reactions. \\

\section{ACKNOWLEDGMENT}
The authors thank Prof. W. Tornow TUNL and Prof. H. Weller TUNL
for valuable discussions.


\begin{thebibliography}{9}                                                                                                %


\bibitem{eji00}
H. Ejiri, Phys. Rep. C 338 265 (2000) and refs. therein.
\bibitem{eji05}
H. Ejiri, J. Phys. Soc. Jap. 74 2101 (2005).
\bibitem{eji06}
H. Ejiri, Czechoslovak J. Phys. 56 459 (2006).
\bibitem{avi08}
F. Avignone, S.R. Elliott and J. Engel, Rev. Mod. Physics 80 481(2008).
\bibitem{eji09}
H. Ejiri, J. Phys. Soc. Japan 78 074201 (2009).
\bibitem{eji10}
H. Ejiri, Progress Particle Nuclear Physics 64 249 (2010).
\bibitem{ver12}
J. Vergados, H. Ejiri and F. Simkovic, Report Progress Physics 75 106301 (2012).
\bibitem{suh98}
J. Suhonen and O. Civitarese, Phys. Rep. 300 123 (1998).
\bibitem{sim11}
F. Simkovic, R. Hodak, A Faessler and P. Vogel, Phys. Rev. C, 83, 015502 (2011).
\bibitem{eji03}
H. Ejiri, Nucl. Instr. Methods A 503 276 (2003).
\bibitem{avi00}
F. Avignone, Workshop Neutr. Nucl. Phys. Stopped $\pi \mu $ Facility (Oak Ridge) (2000).
\bibitem{suh06}
J. Suhonen and M. Kortelainen, Czech J. Phys. 56 519 (2006).
\bibitem{eji13}
 H. Ejiri, et al., J. Phys. Soc. Japan 82 044202 (2013).
\bibitem{aki97}
H. Akimune, et al., Phys. Lett. 394 293 (1997).
\bibitem{thi12}
J.H. Thies, Phys. Rev. C 86 014304 (2012).
\bibitem{thi12a}
J.H. Thies, Phys. Rev. C 86 044309 (2012).
\bibitem{gue11}
C. J. Guess et al., Phys. Rev. C 83, 064318 (2011).
\bibitem{doh08}
H. Dohmann et al., Phys. Rev. C 78, 041602(R) (2008).
\bibitem{fuj67}
J.-I. Fujita, Phys. Lett. B 24 123 (1967).
\bibitem{eji68}
H. Ejiri et al., Phys. Rev. lett. 21 373 (1968).
\bibitem{eji68a}
H. Ejiri, et al., Nucl. Phys. 128 388 (1968).
\bibitem{boh75}
A.Bohr and B. Mottelson, Nuclear Structure II W.A. Benjamin, INC (1975).
\bibitem{don75}
T.W. Donnelly and J.D.Walecka, Electron scattering and nuclear structure, Annual Review Nucl. Science, Vol.25 329 (1975).
\bibitem{eji78}
H. Ejiri and J.I. Fujita, Phys. Rep. C 38 86 (1978) and refs. therein.
\bibitem{eji89}
H. Ejiri and M.J.A. de Voigt, Gamma ray and electron spectroscopy in nuclear physics, Oxford
(1989).
\bibitem{nat12}
National Nuclear Data Center, Brookhaven National Laboratory (2012),
URL http://www.nndc.bnl.gov.
\bibitem{wel09}
H.R. Weller, H.W. Ahmed, H. Gao, W. Tornow, Y.K. Wu, M. Gai, and R. Miskimen, Prog.
Part. Nucl. Phys. 62 257 (2009).
\bibitem{bos13}
M. Boswell, A. Young, and H. Ejiri, presented at a workshop (AIP conference report 2013).
\bibitem{szp13}
B. Szpunar, C. Rangacharyulu, S. Date, and H. Ejiri, Nucl. Instr. Methods, (2013).
\bibitem{miy12}
S. Miyamoto, Private communication (2012).

\end{thebibliography}
\end{document}